\documentclass[reprint,superscriptaddress,
amsmath,amssymb,
aps,
pra,
]{revtex4-2}
\usepackage[paperwidth=210mm,paperheight=297mm,centering,hmargin=1.5cm,vmargin=1.7cm]{geometry}

\usepackage[utf8]{inputenc}

\usepackage{amsmath}
\usepackage{amsthm}
\usepackage{amsfonts}
\usepackage{amssymb}
\usepackage{graphicx}
\usepackage{physics}
\usepackage{siunitx}
\usepackage{color}
\usepackage{multirow}
\usepackage{mathrsfs}
\usepackage{soul}
\usepackage{hhline}
\usepackage{mathtools}
\usepackage{multirow}

\newcommand{\be}{\begin{equation}}
\newcommand{\ee}{\end{equation}}

\newcommand{\db}{\mathsf{D}}

\newcommand{\epscorr}{\varepsilon_{\textrm{cor}}}

\graphicspath{{Images/}}

\begin{document}
\title{Experimental symmetric private information retrieval with measurement-device-independent quantum network}

\author{Chao Wang}
\email{wang.chao@nus.edu.sg}
\affiliation{Department of Electrical \& Computer Engineering, National University of Singapore, Singapore}

\author{Wen Yu Kon}
\affiliation{Department of Electrical \& Computer Engineering, National University of Singapore, Singapore}

\author{Hong Jie Ng}
\affiliation{Department of Electrical \& Computer Engineering, National University of Singapore, Singapore}
 
\author{Charles C.-W. Lim}
\email{charles.lim@nus.edu.sg}
\affiliation{Department of Electrical \& Computer Engineering, National University of Singapore, Singapore}
\affiliation{Centre for Quantum Technologies, National University of Singapore, Singapore}

\begin{abstract}
Secure information retrieval is an essential task in today's highly digitised society. In some applications, it may be necessary that user query's privacy and database content's security are enforced. For these settings, symmetric private information retrieval (SPIR) could be employed, but its implementation is known to be demanding, requiring a private key-exchange network as the base layer.
Here, we report for the first time a realisation of provably-secure SPIR supported by a quantum-secure key-exchange network. 
The SPIR scheme looks at biometric security, offering secure retrieval of 582-byte fingerprint files from a database with 800 entries. 
Our experimental results clearly demonstrate the feasibility of SPIR with quantum secure communications, thereby opening up new possibilities in secure distributed data storage and cloud computing over the future Quantum Internet.
\end{abstract}

\maketitle

\section{Introduction}

Streaming a video on streaming platforms, checking a patient's health records, and verifying one's banking statements -- these are all examples of \emph{information retrieval} (IR), where the goal is to retrieve an entry of interest from an online database. While IR tasks are straightforward to implement, with users sending queries for their desired entries and the data centre responding with the correct information, it becomes challenging when there are privacy concerns. Indeed, from the user's perspective, he/she may not want the data centre to learn about the query of interest for privacy reasons. For instance, the user may not want his/her video preferences to be known by the streaming platform, which can use such information for targeted advertisements. On the other hand, the data managed by the data centre could be sensitive or require long-term security, such as health records or bank account details. As such, these data centres would ideally want other entries of their database to be private from the user.

For tasks requiring both user privacy and database privacy, one can turn to \emph{symmetric private information retrieval} (SPIR), which was first proposed by Gertner et. al.~\cite{Gertner2000}. SPIR guarantees that while performing IR, we have that (1) the data centre cannot learn about the user's query and (2) the user cannot learn more about the database other than the requested information. However, while SPIR can provide strong security guarantees, its implementation is not straightforward. If there is only a single data centre for the user to communicate with, it is known that information-theoretically secure SPIR is impossible even with quantum resources~\cite{Lo1997,Chor1998}. As such, computationally secure SPIR protocols~\cite{Stern1998,Lipmaa2005,Naor2001,Chou2015} and cheat-sensitive quantum private query (QPQ) protocols~\cite{Giovannetti2008,Jakobi2011,Rao2013,Giovannetti2008Proof,Olejnik2011,Li2016} have been proposed. However, computationally secure protocols may be non-ideal for IR tasks with data requiring long-term security since these protocols could be broken with more powerful computers and QPQ does not provide the same strong security guaranteed by SPIR.

To achieve information-theoretic secure SPIR, one can adopt the so-called multi-database scheme proposed by Ref.~\cite{Gertner2000}. In this scheme, the user communicates separately with two or more data centres which holds the same database in order to successfully perform the IR task. If we assume that these data centres are non-communicating, it can be shown that the resulting SPIR scheme is information-theoretically secure~\cite{Gertner2000}. For practical implementation of the proposed scheme, we require additionally that (1) information-theoretic secure communication channels exist between the user and data centres and (2) a random string is securely shared between the data centres. Both requirements can be satisfied with a secure key distribution scheme, since this key can be used directly as a shared random string or together with one-time pad (OTP) encryption for secure communication. However, classical key distribution schemes are not information-theoretic secure, and we have to rely for instance on trusted couriers to deliver the keys, which makes the implementation of SPIR impractical.

To allow for practical implementation of SPIR, we turn to \emph{quantum key distribution} (QKD), an information-theoretically secure method allowing network users to exchange secret keys. By exchanging quantum states and classical communication, QKD allows distant parties to securely generate shared keys which can be later utilised for the SPIR protocol. Since QKD is a relatively mature technology, with commercially available components, extensively-studied security analysis and well-developed post-processing algorithms~\cite{scarani_security_2009,xu_secure_2020}, it provides a basis on which SPIR can be built for practical implementation.

\begin{figure*}[bht!]
    \centering
    \includegraphics[width=0.85\textwidth]{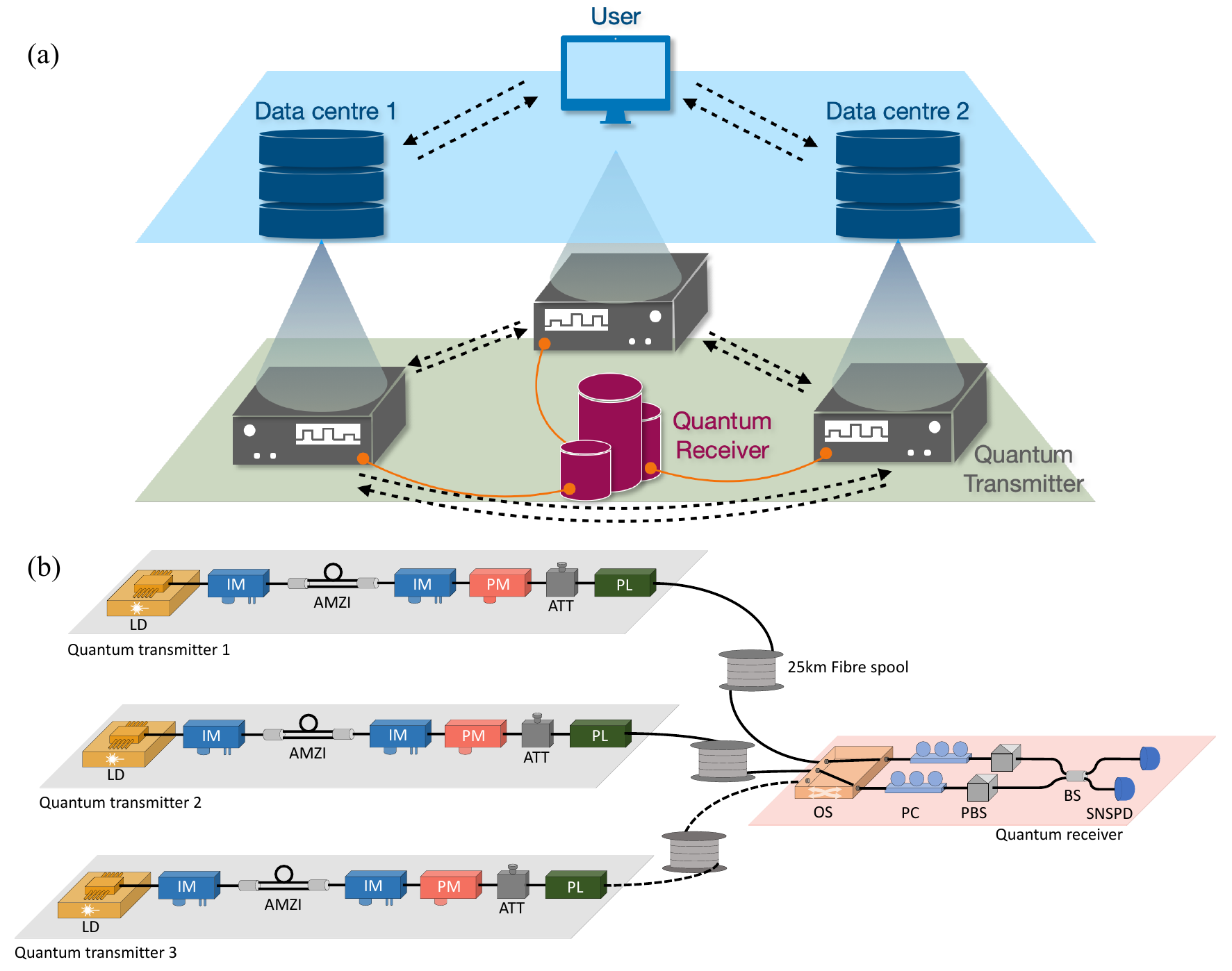}
    \caption{Schematic of our proposed SPIR system. (a) The SPIR system comprises two layers, the QKD layer and the application layer, which operate independently except for the transfer of secret keys. In the QKD layer, quantum transmitters are paired for key distribution, which includes procedures of quantum state preparation, quantum state measurement, and classical post-processing. In the application layer, each party obtains and manages the generated secret keys for the implementation of the SPIR protocol. The black dashed arrows represent the direction of the classical communication, while the orange solid lines represent quantum channels for QKD. (b) Schematic of the MDI QKD implementation. LD: laser diode. IM: intensity modulator. PM: phase modulator. BS: beam splitter. AMZI: asymmetric Mach–Zehnder interferometer. ATT: optical attenuator. PL: optical power limiter. OS: Optical switch. PC: polarisation controller. PBS: polarising beam splitter. SNSPD: superconducting nanowire single-photon detector.}
    \label{SetupFig}
\end{figure*}

The security of SPIR supported with QKD technology has been shown in Ref.~\cite{kon_provably_2021}. Here, we demonstrate, for the first time, a practical implementation of SPIR with a QKD network [see Fig.~\ref{SetupFig} (a)] over a fingerprint database. The specific SPIR protocol utilised is the original protocol proposed by Ref.~\cite{Gertner2000}, using secure keys provided by a \emph{measurement-device-independent} (MDI) QKD protocol~\cite{Lo2012, braunstein_side-channel-free_2012}. The paper is organised as follows. In Sec.~\ref{QKDSPIR}, we briefly introduce the two-database SPIR protocol and the incorporation of QKD to achieve an information-theoretic secure SPIR scheme. In Sec.~\ref{Expt}, we introduce the details of the MDI QKD, present the experimental results, and demonstrate our SPIR scheme on a fingerprint database. In Sec.~\ref{discussion}, we discuss the necessary assumptions and conditions for the proposed scheme. Finally, we conclude in Sec.~\ref{conclusion}.

\section{two-database SPIR with QKD}
\label{QKDSPIR}

\begin{figure*}[hbt!]
    \centering
    \includegraphics[width=0.98\textwidth]{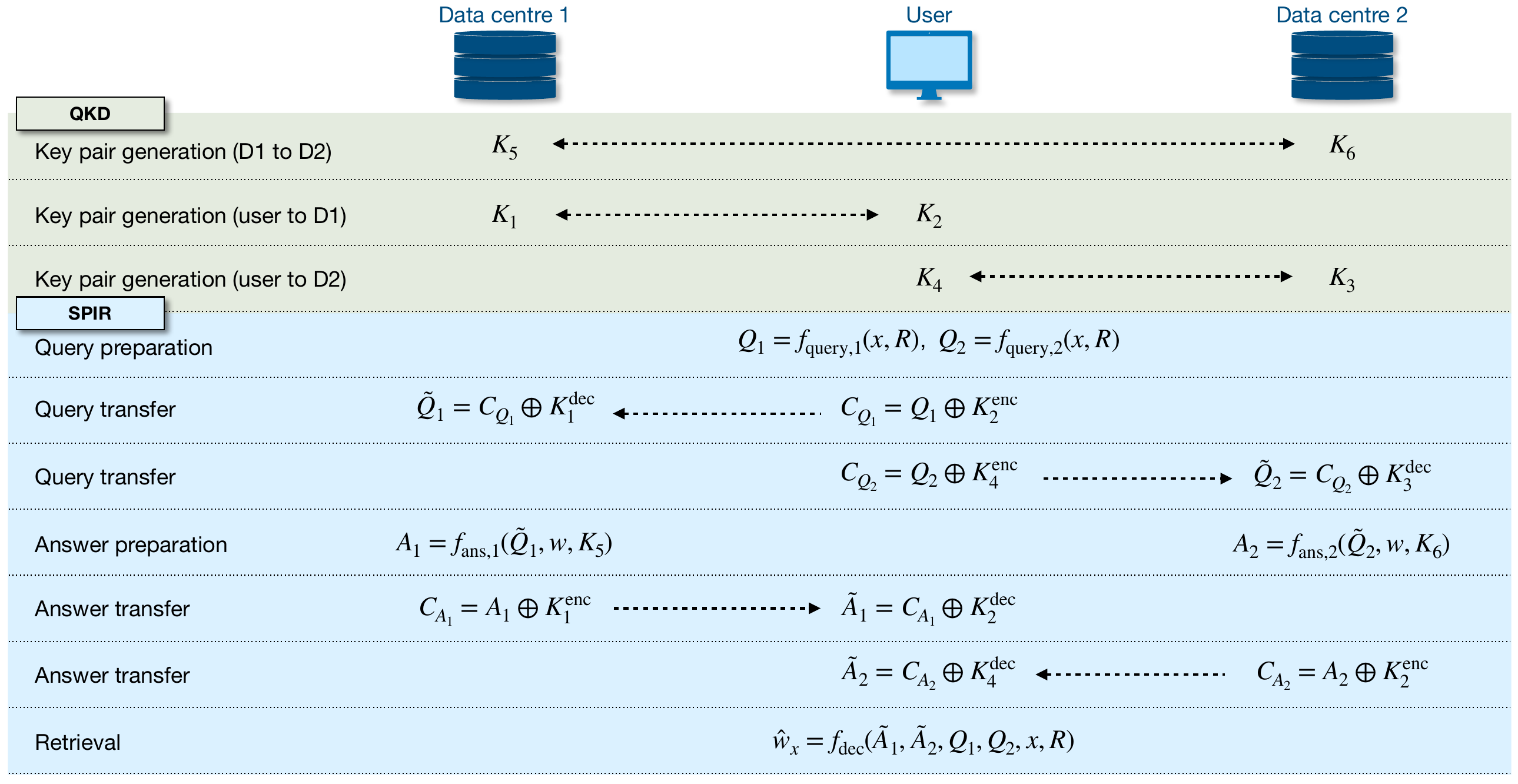}
    \caption{Two-database SPIR protocol, with QKD as the key distribution protocol. }
    \label{workflow}
\end{figure*}

\subsection{Two-Database SPIR}

We consider here an IR scenario, where a user is interested in accessing the $x$-th entry of a database $w$, which contains $n$ different entries $w_i$ ($i\in\{1,...,n\}$), of $L$ bits each. In the corresponding two-database SPIR protocol, the user has to interact with two non-communicating data centres, $\db_1$ and $\db_2$, which each holds a copy of the database $w$. The protocol can be described as follows (also summarised in Fig.~\ref{workflow} with QKD as the key distribution protocol). 

\textbf{Key distribution:} Secret keys are pre-distributed among the various parties in the SPIR protocol. We denote ($K_1$,$K_2$) as the key pair shared between the user and $\db_1$, ($K_3$,$K_4$) for the user and $\db_2$, and ($K_5$,$K_6$) for $\db_1$ and $\db_2$. 

\textbf{Query:} The user prepares queries $Q_i = f_{\rm query,i} (x,R)$, for data centre $\db_i$ ($i\in\{1,2\}$), where $R$ is a random string that is generated by the user locally. Subsequently, the user sends $Q_1$ and $Q_2$ to the respective data centres via OTP encryption with the secret keys $K_2$ and $K_4$, respectively.

\textbf{Answer:} After receiving the encrypted message, $\db_1$ and $\db_2$ first decode the transmitted information using keys $K_1$ and $K_3$. We note that the decrypted queries, $\tilde{Q}_i$, may differ from $Q_i$ if the key pairs are not identical. Thereafter, data centres $\db_1$ and $\db_2$ generate answers $A_1 = f_{\rm ans,1} (\tilde{Q}_1,w,K_5)$ and $A_2 = f_{\rm ans,2} (\tilde{Q}_2,w,K_6)$, respectively. The answers $A_1$ and $A_2$ are then encrypted with keys $K_1$ and $K_3$ and sent to the user. 

\textbf{Retrieval:} Upon receiving the answers from $\db_1$ and $\db_2$, the user decrypt the answers and recovers the desired database entry value with $\hat{w}_x = f_{\rm dec} (\tilde{A}_1,\tilde{A}_2,Q_1,Q_2,x,R)$.

At the end of the SPIR protocol, four conditions should ideally be satisfied. (1) Correctness: The user should correctly recover his desired database entry, i.e. $\hat{w}_x=w_x$. (2) User privacy: The
data centres should not be able to determine the index of the database entry $x$ which the user is interested in. (3) Database privacy: The user should not be able to gain any information beyond a single entry of the database. (4) Protocol secrecy: To protect the security of the data communicated, any external eavesdropper should neither be able to recover $x$ nor any entry of the database $w$.

To achieve the aforementioned security conditions, the use of keys $K_1$ to $K_6$ are essential. Keys $K_1$ to $K_4$ serve as secret keys to encrypt communication between the user and the data centres, preventing any leakage of information to an external eavesdropper or the other data centre that may compromise user privacy and protocol secrecy. Keys $K_5$ and $K_6$ shared between the data centres can be used to mask the answers sent by the data centres to the user, such that the user is only able to retrieve at most one entry of the database, even if the user's action is dishonest~\cite{Gertner2000}. 

In this paper, we focus on having an information-theoretic SPIR protocol, which requires the keys to be distributed with information-theoretic security. 
Having an information-theoretic secure protocol is ideal for data requiring long-term security, such as biometrics and health records, because it hedges against the threat posed by technological advancement. As computing power increases, quantum computers become more powerful, and novel algorithms are developed, many computationally secure protocols are at risk of being broken, which leads to leakage of information to external eavesdroppers. Hence, we require an information-theoretic key distribution protocol to maintain the security of SPIR.

\subsection{SPIR with MDI QKD}

Since information-theoretic secure key distribution is impractical in the classical regime, we propose using QKD to distribute the necessary keys in an information-theoretic secure manner for use in SPIR. The overall SPIR scheme involves running SPIR with the aid of QKD generated keys, as presented in Fig.~\ref{SetupFig} (a). In this scheme, there is a QKD layer responsible for secure key distribution among distant parties with quantum transmitters and receivers. This QKD layer would supply the keys into an application layer upon which the SPIR protocol is implemented. Having this modular structure allows us to not only be flexible in the choice of QKD and SPIR protocols, but also allows other applications, such as secure communication channels, to be built upon the same QKD layer. The formal security proof of the SPIR protocol with QKD keys can be found in Ref.~\cite{kon_provably_2021}. 

In such a SPIR scheme, the final system performance depends on the design for both the SPIR layer and the QKD layer, with two main considerations: practicability and implementation security. 

Practicability is linked to the resources required for implementation. Factors such as the key length and the number of data centres required for SPIR, or the key rate and the topology of the QKD scheme, has to been considered when choosing suitable protocols for the desired application. For instance, the keys required for the SPIR protocol proposed in Ref.~\cite{Gertner2000} for an $n$-entry database with $k$ data centres scales as $O(n^{1/(2k-1)})$. Therefore, for applications with large database sizes, having more data centres can be preferable as it can reduce the key requirements.

Implementation security is closely related to the design and deployment of the QKD layer. Although QKD promises an information-theoretic security for key distribution based on quantum physics, its practical implementation may not be able to fulfil the security conditions perfectly. For example, a finite optical isolation of the quantum transmitter from the outside environment may result in vulnerability to Trojan-horse attacks~\cite{gisin_trojan-horse_2006,vakhitov_large_2001}.

As such, we choose to deploy MDI QKD with decoy states for the QKD layer as it provides a great balance between practicability and implementation security~\cite{Lo2012, Curty2014}. In MDI QKD, each party (user and data centres) holds a quantum transmitter, which needs to be secured. The parties can then communicate via a central quantum receiver, which need not be secure and can be managed by external parties. This gives MDI QKD an appealing feature of immunity against any potential side-channel attacks on the quantum receiver, which is typically regarded as the most vulnerable part in practical QKD implementation~\cite{Lo2012, xu_secure_2020}. As an added advantage, MDI QKD provides a natural star topology, making it suitable for network extension.

The MDI QKD protocol used has a key rate of~\cite{Curty2014}
\begin{equation}
    \begin{split}
        l\leq& n_{0}+n_{1}[1-h(e_{1})]-\text{leak}_{\rm{EC}}\\
        &-\log\frac{8}{\epscorr}-2\log\frac{2}{\varepsilon'\hat{\varepsilon}}-2\log\frac{1}{2\varepsilon_{\rm{PA}}},
    \end{split} \label{eqn:key-length}
\end{equation} 
where $h(x)$ is the binary entropy of $x$, $n_{0}$ is the number of events where either party sends zero photons, $n_{1}$ is the number of events where both parties send one photon each, $e_{1}$ is the error rate of these one-photon events, $\text{leak}_{\rm{EC}}$ is the number of leaked bits from error-correction, and the various $\varepsilon$ values are security parameters.

For the SPIR layer, we consider the two-database SPIR protocol proposed in Ref.~\cite{Gertner2000} (detailed also in Appendix B of Ref.~\cite{kon_provably_2021}). For a database with $n$ entries of length $L$, the protocol requires $[7L+3\lceil\log m\rceil+(3+3L)m]$ bits of key for secure communication between the user and each data centre, and $(9Lm+10L)$ bits of key for use as shared random bits between the data centres, where $m=\lceil n^{1/3}\rceil$.

\section{Simulation and Experimental results}
\label{Expt}

\subsection{Experimental details of MDI QKD}
\label{exptdetail}

The experimental setup of the MDI QKD is shown in Fig.~\ref{SetupFig} (b). The quantum transmitter held by each party consists of a laser source section and a quantum state preparation section. In the laser source section, a distributed feedback laser diode is operated in the gain-switching mode to generate laser pulses with a repetition rate of 125 MHz. This allows each optical pulse to inherit an intrinsically random and independent phase~\cite{kobayashi_evaluation_2014,yuan_robust_2014} required for decoy-state analysis~\cite{tang_source_2013}. An intensity modulator~(IM) is used for further pulse carving, which generates optical pulses with 220~ps width. 
In the quantum state preparation section, the phase randomised optical pulses are split into earlier and later time-bins by an asymmetric Mach-Zehnder interferometer. Thereafter, the pulses are modulated by an IM, a phase modulator (PM) and optical attenuators to generate time-bin phase-encoded quantum states: $\ket{0}=\ket{e}_{\mu_j}$, $\ket{1}=\ket{l}_{\mu_j}$, $\ket{2} = (\ket{e}_{\mu_j}+\ket{l}_{\mu_j})/\sqrt{2}$, $\ket{3} = (\ket{e}_{\mu_j}-\ket{l}_{\mu_j})/\sqrt{2}$, where $\ket{e}$ and $\ket{l}$ represents the early and late time-bin temporal modes, and $\mu_j$, $j\in\{1,2,3\}$, represents three different intensities for the purpose of decoy-state analysis. Finally, an optical power limiter~\cite{zhang_securing_2020} (or optical isolators~\cite{lucamarini_practical_2015}) is used to limit the information leakage from the transmitter to the outside environment. The central wavelength of the laser diodes are fine-tuned with a precision of around 0.1~pm (corresponding to a frequency uncertainty of 12.5~MHz), which guarantees the indistinguishability in the spectral mode of the two quantum states. Moreover, as the quantum states are required to arrive simultaneously at the receiver, the laser pulse generation and signal modulations in each transmitter are all synchronised to the same master clock, with a timing delay configuration precision of 10 ps. 

After going through a \SI{25}{\kilo\meter} spooled optical fibre, the quantum states from each quantum transmitter arrive at the quantum receiver. In the quantum receiver, an optical switch works in a time-division multiplexing way to connect two of the three parties to the quantum receiver. After successfully linking two legitimate parties to the quantum receiver, the fibre optical polarisation controllers and polarising beam splitters in each path calibrate the state of polarisation of the incoming photons. Thereafter, a 50:50 fibre beam splitter and two superconducting nanowire single-photon detectors perform Bell-state measurement (BSM) on the input quantum states. Including the insertion losses of optical components and fibre connectors, the final effective quantum efficiency of the measurement devices is 70.73\% on average. 

After the BSM, the quantum receiver publicly announces the measurement results. The paired transmitters then perform the necessary data processing and negotiation over an authenticated classical communication channel, including basis sifting, error correction and privacy amplification, etc., to obtain the final identical secure keys. 

After calibrating all the degrees of freedom of the independent quantum transmitters, the Hong-Ou-Mandel interference visibility is measured to be 0.48 ($\pm$0.015). The slight deviation from the theoretical value of 0.5 with coherent state inputs and perfect mode overlapping indicates a good indistinguishability of the generated quantum states, which is a prerequisite for high efficiency BSM and determines the performance of the MDI QKD system.

\begin{figure}[hbt!]
    \centering
    \includegraphics[width=0.48\textwidth]{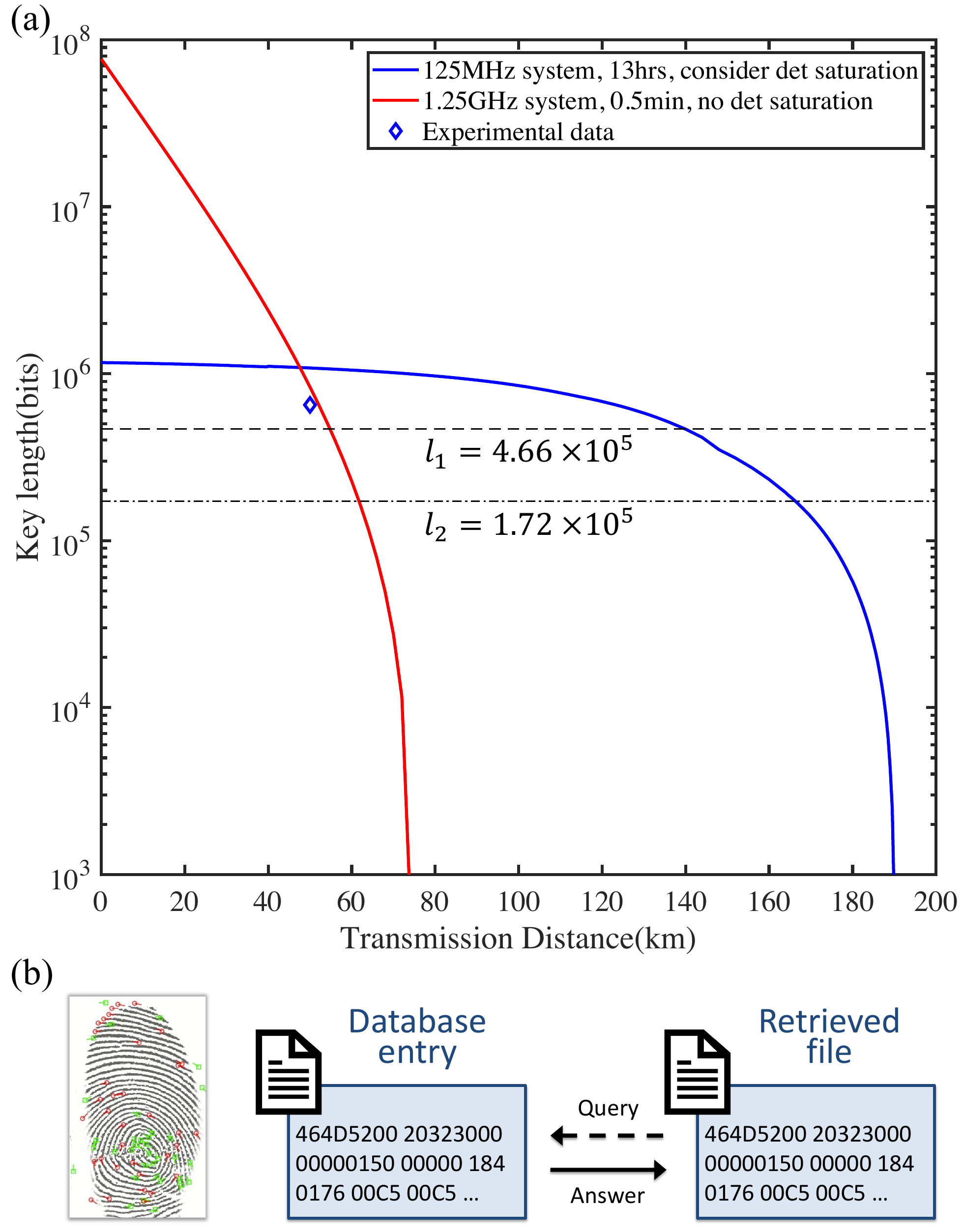}
    \caption{Simulation and experimental result of the MDI QKD system. (a) The blue curve is the simulated secure key length with $N = 5.85\times 10^{13}$ signal pulses (13 hours operation of \SI{125}{\mega\hertz} system) and with detector saturation under consideration. The red curve is the simulated secure key length with $N = 3.75\times 10^{10}$ signal pulses (0.5 minute operation of \SI{1.25}{\giga\hertz} system) and with no intensity limitation for quantum state preparation. The dotted lines indicate the number of keys required for the fingerprint database (\SI{4.66e5}{} shared bits of randomness between data centres and \SI{1.72e5}{} shared secret bits between the user and each data centre). (b) Demonstration of the fingerprint retrieval. 
    }
    \label{PlotFig}
\end{figure}

\subsection{SPIR demonstration on fingerprint database}
\label{demonstration}

Here, we demonstrate the SPIR scheme with MDI QKD keys using a fingerprint minutiae database (containing only key features of the fingerprint) stored in the ISO 19794-2 standard format~\cite{ISO_Fingerprint}. The database chosen is DB1A of the Fingerprint Verification Competition 2002~\cite{Maltoni2009}, which is converted into minutiae data by Kayaoglu et. al.~\cite{Kayaoglu2013}. It contains 800 entries ($n=800$) and the maximum file size is 582 bytes ($L=4656$). As such, \SI{1.72e5}{} bits of secret keys is required between the user and each data centre and \SI{4.66e5}{} shared random bits is required between the two data centres.

To verify the feasibility of the application, we first study its performance using a key rate simulation~\cite{yuan_interference_2014,ma_alternative_2012,Curty2014,wang_realistic_2017}. Using realistic parameters obtained from the MDI QKD system, with security parameters $\varepsilon_{corr}=$\SI{e-15}{} and $\varepsilon_{sec}=$\SI{e-10}{}, we optimise $l$ in Eqn.~\eqref{eqn:key-length} over the intensities $\mu_1$, $\mu_2$, $\mu_3$, the probability of choosing an intensity and basis combination, the number of bits used for parameter estimation, and various security parameters. The performance of two cases are simulated. In the first case, we study our current MDI QKD system with a working frequency of \SI{125}{MHz} and a run time of 13 hours, generating $N = 5.85\times 10^{12}$ signal pulses in total. We also take the SNSPD counting rate saturation into consideration, which limits the value of $\mu_1$ to a maximum value that varies with transmission distance. In the second case, we study our current MDI QKD system, but set at a working frequency of \SI{1.25}{GHz}~\cite{wei_high-speed_2020} and a run time of 0.5 minutes, generating $N = 3.75\times 10^{10}$ signal pulses. In addition, high counting rate single-photon detectors are assumed for quantum state measurement~\cite{dauler_125-gbits_2006,chen_twin-field_2021}. In this case, the photon counting saturation issue is negligible, and there is no constraint on the value of $\mu_1$. The simulation results are shown as the blue and red curve in Fig.~\ref{PlotFig} (a) for the two cases. In both cases, the final key length generated is sufficient to meet the key requirement of the SPIR protocol (labelled by dotted lines) at a \SI{50}{km} transmission distance.

We use the optimised parameters in the first case for our MDI QKD system to perform the experiment. After the quantum state preparation and quantum state measurement stated in Sec.~\ref{exptdetail}, we first obtain raw keys as well as all the necessary statistics for key rate calculation. Subsequently, we perform post-processing on the raw key, including basis sifting, error correction, and privacy amplification to obtain the final secure keys required for SPIR. 

After basis sifting, 10.34\% of the sifted keys are used for parameter estimation, where it was found that the average bit error rate when transmitting states in the Z basis ($\ket{0}$ and $\ket{1}$) is 0.83\%. Error correction is performed on the remaining sifted keys by using symmetric blind low-density parity-check (LDPC) code~\cite{kiktenko_symmetric_2017}. This error correction code achieves an average correction efficiency of $f_{\rm EC} = 1.41$. We note here that a better efficiency performance can be achieved by applying some improved LDPC schemes~\cite{elkouss_efficient_2009}, or using some interactive protocols, e.g. \emph{Cascade}, which provide higher efficiency in the low error rate region~\cite{elkouss_efficient_2009,kiktenko_symmetric_2017}. However, for the latter case, a finite-key analysis with two-way protocols is desired for a rigorous security proof~\cite{tomamichel_fundamental_2017}. Subsequently, the corrected keys undergo privacy amplification via Toeplitz hashing~\cite{krawczyk_lfsr-based_1994} accelerated by fast Fourier transform to generate final keys that are secret and uniformly distributed. 

After post-processing, $l = 6.50\times 10^{5}$ bits of final secure keys are extracted. Here, for our proof-of-concept demonstration, only two quantum transmitters are actually implemented and the same pair of generated keys are re-used for all three QKD links. Finally, we implement the SPIR protocol, and successfully retrieve a target fingerprint file (4656 bits) from the database. 

\section{Discussion}
\label{discussion}

We have demonstrated experimentally that the overall SPIR scheme is feasible. However, it is important to note the necessary assumptions and conditions required for a proper implementation of the protocol.

Firstly, to ensure the security of the SPIR protocol, we assume that the data centres are non-communicating~\cite{Gertner2000,kon_provably_2021}. If the two data centres are allowed to communicate, they are then able to behave like a single entity, which renders the SPIR protocol insecure. To enforce this assumption in practice, we could expect that administrative network management and access controls be utilised to prevent unauthorised communications~\cite{robling_denning_cryptography_1982,kim_fundamentals_2016}. 

Secondly, we assume that the QKD system operates independently from the application layer. More specifically, none of the parties involved in SPIR (user or data centres), should be allowed access to the internal components of the quantum transmitter, or control its operations. To enforce this, one can reasonably imagine that the quantum transmitters are properly sealed and shielded by the service provider. 

Finally, we have to also consider additional assumptions related to the QKD implementation security. For instance, in MDI QKD, the quantum transmitters are assumed to be secure and inaccessible to the eavesdropper. As such, any attacks from the quantum communication channel should be kept under control. Fortunately, such problems have been studied as source-related attacks in the practical QKD security analysis, such as the Trojan-horse attack~\cite{vakhitov_large_2001,gisin_trojan-horse_2006}, laser seeding attack~\cite{sun_effect_2015,huang_laser-seeding_2019} and laser damage attack~\cite{huang_laser-damage_2020}. Since the eavesdropping light is injected into the quantum transmitter via the quantum channel, a countermeasure based on optical power control can be expected, such as optical power limiter~\cite{zhang_securing_2020} and optical fibre isolators~\cite{lucamarini_practical_2015}.

\section{Conclusion}
\label{conclusion}

In this paper, we experimentally demonstrated a two-database SPIR scheme utilising keys from a MDI QKD system on a fingerprint database with 800 entries and a maximum of 582 bytes per entry. In this two-layered scheme, the QKD layer generates the necessary keys for the SPIR protocol in the application layer, where they are used for secure communication and as shared random bit strings. This allows the overall SPIR scheme to be information-theoretic secure, thus satisfying the strong security guarantees that certain IR problems may require, especially ones that involves sensitive data or data requiring long-term security. 
Our proposed scheme, along with its demonstration here, thus illustrates the feasibility of the practical implementation of SPIR for tacking IR problems.

\section{Acknowledgements}
\label{acknowledgements}
This research is supported by the National Research Foundation (NRF) Singapore, under its NRF Fellowship programme (NRFF11-2019-0001) and Quantum Engineering Programme 1.0 projects (QEP-P2, QEP-P3, and QEP-P8)

\bibliography{QKDSPIRManu}

\begin{thebibliography}{44}%
\makeatletter
\providecommand \@ifxundefined [1]{%
 \@ifx{#1\undefined}
}%
\providecommand \@ifnum [1]{%
 \ifnum #1\expandafter \@firstoftwo
 \else \expandafter \@secondoftwo
 \fi
}%
\providecommand \@ifx [1]{%
 \ifx #1\expandafter \@firstoftwo
 \else \expandafter \@secondoftwo
 \fi
}%
\providecommand \natexlab [1]{#1}%
\providecommand \enquote  [1]{``#1''}%
\providecommand \bibnamefont  [1]{#1}%
\providecommand \bibfnamefont [1]{#1}%
\providecommand \citenamefont [1]{#1}%
\providecommand \href@noop [0]{\@secondoftwo}%
\providecommand \href [0]{\begingroup \@sanitize@url \@href}%
\providecommand \@href[1]{\@@startlink{#1}\@@href}%
\providecommand \@@href[1]{\endgroup#1\@@endlink}%
\providecommand \@sanitize@url [0]{\catcode `\\12\catcode `\$12\catcode
  `\&12\catcode `\#12\catcode `\^12\catcode `\_12\catcode `\%12\relax}%
\providecommand \@@startlink[1]{}%
\providecommand \@@endlink[0]{}%
\providecommand \url  [0]{\begingroup\@sanitize@url \@url }%
\providecommand \@url [1]{\endgroup\@href {#1}{\urlprefix }}%
\providecommand \urlprefix  [0]{URL }%
\providecommand \Eprint [0]{\href }%
\providecommand \doibase [0]{http://dx.doi.org/}%
\providecommand \selectlanguage [0]{\@gobble}%
\providecommand \bibinfo  [0]{\@secondoftwo}%
\providecommand \bibfield  [0]{\@secondoftwo}%
\providecommand \translation [1]{[#1]}%
\providecommand \BibitemOpen [0]{}%
\providecommand \bibitemStop [0]{}%
\providecommand \bibitemNoStop [0]{.\EOS\space}%
\providecommand \EOS [0]{\spacefactor3000\relax}%
\providecommand \BibitemShut  [1]{\csname bibitem#1\endcsname}%
\let\auto@bib@innerbib\@empty
\bibitem [{\citenamefont {Gertner}\ \emph {et~al.}(2000)\citenamefont
  {Gertner}, \citenamefont {Ishai}, \citenamefont {Kushilevitz},\ and\
  \citenamefont {Malkin}}]{Gertner2000}%
  \BibitemOpen
  \bibfield  {author} {\bibinfo {author} {\bibfnamefont {Y.}~\bibnamefont
  {Gertner}}, \bibinfo {author} {\bibfnamefont {Y.}~\bibnamefont {Ishai}},
  \bibinfo {author} {\bibfnamefont {E.}~\bibnamefont {Kushilevitz}}, \ and\
  \bibinfo {author} {\bibfnamefont {T.}~\bibnamefont {Malkin}},\ }\href
  {\doibase 10.1006/jcss.1999.1689} {\bibfield  {journal} {\bibinfo  {journal}
  {J. Comput. Syst. Sci.}\ }\textbf {\bibinfo {volume} {60}},\ \bibinfo {pages}
  {592} (\bibinfo {year} {2000})}\BibitemShut {NoStop}%
\bibitem [{\citenamefont {Lo}(1997)}]{Lo1997}%
  \BibitemOpen
  \bibfield  {author} {\bibinfo {author} {\bibfnamefont {H.-K.}\ \bibnamefont
  {Lo}},\ }\href {\doibase 10.1103/PhysRevA.56.1154} {\bibfield  {journal}
  {\bibinfo  {journal} {Phys. Rev. A}\ }\textbf {\bibinfo {volume} {56}},\
  \bibinfo {pages} {1154} (\bibinfo {year} {1997})}\BibitemShut {NoStop}%
\bibitem [{\citenamefont {Chor}\ \emph {et~al.}(1998)\citenamefont {Chor},
  \citenamefont {Kushilevitz}, \citenamefont {Goldreich},\ and\ \citenamefont
  {Sudan}}]{Chor1998}%
  \BibitemOpen
  \bibfield  {author} {\bibinfo {author} {\bibfnamefont {B.}~\bibnamefont
  {Chor}}, \bibinfo {author} {\bibfnamefont {E.}~\bibnamefont {Kushilevitz}},
  \bibinfo {author} {\bibfnamefont {O.}~\bibnamefont {Goldreich}}, \ and\
  \bibinfo {author} {\bibfnamefont {M.}~\bibnamefont {Sudan}},\ }\href
  {\doibase 10.1145/293347.293350} {\bibfield  {journal} {\bibinfo  {journal}
  {J. ACM}\ }\textbf {\bibinfo {volume} {45}},\ \bibinfo {pages} {965–981}
  (\bibinfo {year} {1998})}\BibitemShut {NoStop}%
\bibitem [{\citenamefont {Stern}(1998)}]{Stern1998}%
  \BibitemOpen
  \bibfield  {author} {\bibinfo {author} {\bibfnamefont {J.~P.}\ \bibnamefont
  {Stern}},\ }in\ \href@noop {} {\emph {\bibinfo {booktitle} {Advances in
  Cryptology --- ASIACRYPT'98}}},\ \bibinfo {editor} {edited by\ \bibinfo
  {editor} {\bibfnamefont {K.}~\bibnamefont {Ohta}}\ and\ \bibinfo {editor}
  {\bibfnamefont {D.}~\bibnamefont {Pei}}}\ (\bibinfo  {publisher} {Springer
  Berlin Heidelberg},\ \bibinfo {address} {Berlin, Heidelberg},\ \bibinfo
  {year} {1998})\ pp.\ \bibinfo {pages} {357--371}\BibitemShut {NoStop}%
\bibitem [{\citenamefont {Lipmaa}(2005)}]{Lipmaa2005}%
  \BibitemOpen
  \bibfield  {author} {\bibinfo {author} {\bibfnamefont {H.}~\bibnamefont
  {Lipmaa}},\ }in\ \href@noop {} {\emph {\bibinfo {booktitle} {Information
  Security}}},\ \bibinfo {editor} {edited by\ \bibinfo {editor} {\bibfnamefont
  {J.}~\bibnamefont {Zhou}}, \bibinfo {editor} {\bibfnamefont {J.}~\bibnamefont
  {Lopez}}, \bibinfo {editor} {\bibfnamefont {R.~H.}\ \bibnamefont {Deng}}, \
  and\ \bibinfo {editor} {\bibfnamefont {F.}~\bibnamefont {Bao}}}\ (\bibinfo
  {publisher} {Springer Berlin Heidelberg},\ \bibinfo {address} {Berlin,
  Heidelberg},\ \bibinfo {year} {2005})\ pp.\ \bibinfo {pages}
  {314--328}\BibitemShut {NoStop}%
\bibitem [{\citenamefont {Naor}\ and\ \citenamefont {Pinkas}(2001)}]{Naor2001}%
  \BibitemOpen
  \bibfield  {author} {\bibinfo {author} {\bibfnamefont {M.}~\bibnamefont
  {Naor}}\ and\ \bibinfo {author} {\bibfnamefont {B.}~\bibnamefont {Pinkas}},\
  }in\ \href@noop {} {\emph {\bibinfo {booktitle} {Proceedings of the twelfth
  annual {ACM}-{SIAM} symposium on {Discrete} algorithms}}},\ \bibinfo {series
  and number} {{SODA} '01}\ (\bibinfo  {publisher} {Society for Industrial and
  Applied Mathematics},\ \bibinfo {address} {Washington, D.C., USA},\ \bibinfo
  {year} {2001})\ pp.\ \bibinfo {pages} {448--457}\BibitemShut {NoStop}%
\bibitem [{\citenamefont {Chou}\ and\ \citenamefont
  {Orlandi}(2015)}]{Chou2015}%
  \BibitemOpen
  \bibfield  {author} {\bibinfo {author} {\bibfnamefont {T.}~\bibnamefont
  {Chou}}\ and\ \bibinfo {author} {\bibfnamefont {C.}~\bibnamefont {Orlandi}},\
  }in\ \href {\doibase 10.1007/978-3-319-22174-8_3} {\emph {\bibinfo
  {booktitle} {Progress in {Cryptology} -- {LATINCRYPT} 2015}}},\ \bibinfo
  {series and number} {Lecture {Notes} in {Computer} {Science}},\ \bibinfo
  {editor} {edited by\ \bibinfo {editor} {\bibfnamefont {K.}~\bibnamefont
  {Lauter}}\ and\ \bibinfo {editor} {\bibfnamefont {F.}~\bibnamefont
  {Rodríguez-Henríquez}}}\ (\bibinfo  {publisher} {Springer International
  Publishing},\ \bibinfo {address} {Cham},\ \bibinfo {year} {2015})\ pp.\
  \bibinfo {pages} {40--58}\BibitemShut {NoStop}%
\bibitem [{\citenamefont {Giovannetti}\ \emph {et~al.}(2008)\citenamefont
  {Giovannetti}, \citenamefont {Lloyd},\ and\ \citenamefont
  {Maccone}}]{Giovannetti2008}%
  \BibitemOpen
  \bibfield  {author} {\bibinfo {author} {\bibfnamefont {V.}~\bibnamefont
  {Giovannetti}}, \bibinfo {author} {\bibfnamefont {S.}~\bibnamefont {Lloyd}},
  \ and\ \bibinfo {author} {\bibfnamefont {L.}~\bibnamefont {Maccone}},\ }\href
  {\doibase 10.1103/PhysRevLett.100.230502} {\bibfield  {journal} {\bibinfo
  {journal} {Phys. Rev. Lett.}\ }\textbf {\bibinfo {volume} {100}},\ \bibinfo
  {pages} {230502} (\bibinfo {year} {2008})}\BibitemShut {NoStop}%
\bibitem [{\citenamefont {Jakobi}\ \emph {et~al.}(2011)\citenamefont {Jakobi},
  \citenamefont {Simon}, \citenamefont {Gisin}, \citenamefont {Bancal},
  \citenamefont {Branciard}, \citenamefont {Walenta},\ and\ \citenamefont
  {Zbinden}}]{Jakobi2011}%
  \BibitemOpen
  \bibfield  {author} {\bibinfo {author} {\bibfnamefont {M.}~\bibnamefont
  {Jakobi}}, \bibinfo {author} {\bibfnamefont {C.}~\bibnamefont {Simon}},
  \bibinfo {author} {\bibfnamefont {N.}~\bibnamefont {Gisin}}, \bibinfo
  {author} {\bibfnamefont {J.-D.}\ \bibnamefont {Bancal}}, \bibinfo {author}
  {\bibfnamefont {C.}~\bibnamefont {Branciard}}, \bibinfo {author}
  {\bibfnamefont {N.}~\bibnamefont {Walenta}}, \ and\ \bibinfo {author}
  {\bibfnamefont {H.}~\bibnamefont {Zbinden}},\ }\href {\doibase
  10.1103/PhysRevA.83.022301} {\bibfield  {journal} {\bibinfo  {journal} {Phys.
  Rev. A}\ }\textbf {\bibinfo {volume} {83}},\ \bibinfo {pages} {022301}
  (\bibinfo {year} {2011})}\BibitemShut {NoStop}%
\bibitem [{\citenamefont {Panduranga~Rao}\ and\ \citenamefont
  {Jakobi}(2013)}]{Rao2013}%
  \BibitemOpen
  \bibfield  {author} {\bibinfo {author} {\bibfnamefont {M.~V.}\ \bibnamefont
  {Panduranga~Rao}}\ and\ \bibinfo {author} {\bibfnamefont {M.}~\bibnamefont
  {Jakobi}},\ }\href {\doibase 10.1103/PhysRevA.87.012331} {\bibfield
  {journal} {\bibinfo  {journal} {Phys. Rev. A}\ }\textbf {\bibinfo {volume}
  {87}},\ \bibinfo {pages} {012331} (\bibinfo {year} {2013})}\BibitemShut
  {NoStop}%
\bibitem [{\citenamefont {{Giovannetti}}\ \emph {et~al.}(2010)\citenamefont
  {{Giovannetti}}, \citenamefont {{Lloyd}},\ and\ \citenamefont
  {{Maccone}}}]{Giovannetti2008Proof}%
  \BibitemOpen
  \bibfield  {author} {\bibinfo {author} {\bibfnamefont {V.}~\bibnamefont
  {{Giovannetti}}}, \bibinfo {author} {\bibfnamefont {S.}~\bibnamefont
  {{Lloyd}}}, \ and\ \bibinfo {author} {\bibfnamefont {L.}~\bibnamefont
  {{Maccone}}},\ }\href {\doibase 10.1109/TIT.2010.2048446} {\bibfield
  {journal} {\bibinfo  {journal} {IEEE Transactions on Information Theory}\
  }\textbf {\bibinfo {volume} {56}},\ \bibinfo {pages} {3465} (\bibinfo {year}
  {2010})}\BibitemShut {NoStop}%
\bibitem [{\citenamefont {Olejnik}(2011)}]{Olejnik2011}%
  \BibitemOpen
  \bibfield  {author} {\bibinfo {author} {\bibfnamefont {L.}~\bibnamefont
  {Olejnik}},\ }\href {\doibase 10.1103/PhysRevA.84.022313} {\bibfield
  {journal} {\bibinfo  {journal} {Phys. Rev. A}\ }\textbf {\bibinfo {volume}
  {84}},\ \bibinfo {pages} {022313} (\bibinfo {year} {2011})}\BibitemShut
  {NoStop}%
\bibitem [{\citenamefont {Li}\ \emph {et~al.}(2016)\citenamefont {Li},
  \citenamefont {Yang}, \citenamefont {Chen}, \citenamefont {Zhou},\ and\
  \citenamefont {Shi}}]{Li2016}%
  \BibitemOpen
  \bibfield  {author} {\bibinfo {author} {\bibfnamefont {J.}~\bibnamefont
  {Li}}, \bibinfo {author} {\bibfnamefont {Y.-G.}\ \bibnamefont {Yang}},
  \bibinfo {author} {\bibfnamefont {X.-B.}\ \bibnamefont {Chen}}, \bibinfo
  {author} {\bibfnamefont {Y.-H.}\ \bibnamefont {Zhou}}, \ and\ \bibinfo
  {author} {\bibfnamefont {W.-M.}\ \bibnamefont {Shi}},\ }\href {\doibase
  10.1038/srep31738} {\bibfield  {journal} {\bibinfo  {journal} {Sci. Rep.}\
  }\textbf {\bibinfo {volume} {6}},\ \bibinfo {pages} {31738} (\bibinfo {year}
  {2016})}\BibitemShut {NoStop}%
\bibitem [{\citenamefont {Scarani}\ \emph {et~al.}(2009)\citenamefont
  {Scarani}, \citenamefont {Bechmann-Pasquinucci}, \citenamefont {Cerf},
  \citenamefont {Dusek}, \citenamefont {Lutkenhaus},\ and\ \citenamefont
  {Peev}}]{scarani_security_2009}%
  \BibitemOpen
  \bibfield  {author} {\bibinfo {author} {\bibfnamefont {V.}~\bibnamefont
  {Scarani}}, \bibinfo {author} {\bibfnamefont {H.}~\bibnamefont
  {Bechmann-Pasquinucci}}, \bibinfo {author} {\bibfnamefont {N.~J.}\
  \bibnamefont {Cerf}}, \bibinfo {author} {\bibfnamefont {M.}~\bibnamefont
  {Dusek}}, \bibinfo {author} {\bibfnamefont {N.}~\bibnamefont {Lutkenhaus}}, \
  and\ \bibinfo {author} {\bibfnamefont {M.}~\bibnamefont {Peev}},\ }\href@noop
  {} {\bibfield  {journal} {\bibinfo  {journal} {Rev. Mod. Phys.}\ }\textbf
  {\bibinfo {volume} {81}},\ \bibinfo {pages} {1301} (\bibinfo {year}
  {2009})}\BibitemShut {NoStop}%
\bibitem [{\citenamefont {Xu}\ \emph {et~al.}(2020)\citenamefont {Xu},
  \citenamefont {Ma}, \citenamefont {Zhang}, \citenamefont {Lo},\ and\
  \citenamefont {Pan}}]{xu_secure_2020}%
  \BibitemOpen
  \bibfield  {author} {\bibinfo {author} {\bibfnamefont {F.}~\bibnamefont
  {Xu}}, \bibinfo {author} {\bibfnamefont {X.}~\bibnamefont {Ma}}, \bibinfo
  {author} {\bibfnamefont {Q.}~\bibnamefont {Zhang}}, \bibinfo {author}
  {\bibfnamefont {H.-K.}\ \bibnamefont {Lo}}, \ and\ \bibinfo {author}
  {\bibfnamefont {J.-W.}\ \bibnamefont {Pan}},\ }\href@noop {} {\bibfield
  {journal} {\bibinfo  {journal} {Rev. Mod. Phys.}\ }\textbf {\bibinfo {volume}
  {92}},\ \bibinfo {pages} {025002} (\bibinfo {year} {2020})}\BibitemShut
  {NoStop}%
\bibitem [{\citenamefont {Kon}\ and\ \citenamefont
  {Lim}(2021)}]{kon_provably_2021}%
  \BibitemOpen
  \bibfield  {author} {\bibinfo {author} {\bibfnamefont {W.~Y.}\ \bibnamefont
  {Kon}}\ and\ \bibinfo {author} {\bibfnamefont {C.~C.~W.}\ \bibnamefont
  {Lim}},\ }\href@noop {} {\bibfield  {journal} {\bibinfo  {journal} {Entropy}\
  }\textbf {\bibinfo {volume} {23}},\ \bibinfo {pages} {54} (\bibinfo {year}
  {2021})}\BibitemShut {NoStop}%
\bibitem [{\citenamefont {Lo}\ \emph {et~al.}(2012)\citenamefont {Lo},
  \citenamefont {Curty},\ and\ \citenamefont {Qi}}]{Lo2012}%
  \BibitemOpen
  \bibfield  {author} {\bibinfo {author} {\bibfnamefont {H.-K.}\ \bibnamefont
  {Lo}}, \bibinfo {author} {\bibfnamefont {M.}~\bibnamefont {Curty}}, \ and\
  \bibinfo {author} {\bibfnamefont {B.}~\bibnamefont {Qi}},\ }\href@noop {}
  {\bibfield  {journal} {\bibinfo  {journal} {Phys. Rev. Lett.}\ }\textbf
  {\bibinfo {volume} {108}},\ \bibinfo {pages} {130503} (\bibinfo {year}
  {2012})}\BibitemShut {NoStop}%
\bibitem [{\citenamefont {Braunstein}\ and\ \citenamefont
  {Pirandola}(2012)}]{braunstein_side-channel-free_2012}%
  \BibitemOpen
  \bibfield  {author} {\bibinfo {author} {\bibfnamefont {S.~L.}\ \bibnamefont
  {Braunstein}}\ and\ \bibinfo {author} {\bibfnamefont {S.}~\bibnamefont
  {Pirandola}},\ }\href@noop {} {\bibfield  {journal} {\bibinfo  {journal}
  {Phys. Rev. Lett.}\ }\textbf {\bibinfo {volume} {108}},\ \bibinfo {pages}
  {130502} (\bibinfo {year} {2012})}\BibitemShut {NoStop}%
\bibitem [{\citenamefont {Gisin}\ \emph {et~al.}(2006)\citenamefont {Gisin},
  \citenamefont {Fasel}, \citenamefont {Kraus}, \citenamefont {Zbinden},\ and\
  \citenamefont {Ribordy}}]{gisin_trojan-horse_2006}%
  \BibitemOpen
  \bibfield  {author} {\bibinfo {author} {\bibfnamefont {N.}~\bibnamefont
  {Gisin}}, \bibinfo {author} {\bibfnamefont {S.}~\bibnamefont {Fasel}},
  \bibinfo {author} {\bibfnamefont {B.}~\bibnamefont {Kraus}}, \bibinfo
  {author} {\bibfnamefont {H.}~\bibnamefont {Zbinden}}, \ and\ \bibinfo
  {author} {\bibfnamefont {G.}~\bibnamefont {Ribordy}},\ }\href@noop {}
  {\bibfield  {journal} {\bibinfo  {journal} {Phys. Rev. A}\ }\textbf {\bibinfo
  {volume} {73}},\ \bibinfo {pages} {022320} (\bibinfo {year}
  {2006})}\BibitemShut {NoStop}%
\bibitem [{\citenamefont {Vakhitov}\ \emph {et~al.}(2001)\citenamefont
  {Vakhitov}, \citenamefont {Makarov},\ and\ \citenamefont
  {Hjelme}}]{vakhitov_large_2001}%
  \BibitemOpen
  \bibfield  {author} {\bibinfo {author} {\bibfnamefont {A.}~\bibnamefont
  {Vakhitov}}, \bibinfo {author} {\bibfnamefont {V.}~\bibnamefont {Makarov}}, \
  and\ \bibinfo {author} {\bibfnamefont {D.~R.}\ \bibnamefont {Hjelme}},\
  }\href@noop {} {\bibfield  {journal} {\bibinfo  {journal} {J. Mod. Opt.}\
  }\textbf {\bibinfo {volume} {48}},\ \bibinfo {pages} {2023} (\bibinfo {year}
  {2001})}\BibitemShut {NoStop}%
\bibitem [{\citenamefont {Curty}\ \emph {et~al.}(2014)\citenamefont {Curty},
  \citenamefont {Xu}, \citenamefont {Lim}, \citenamefont {Tamaki},\ and\
  \citenamefont {Lo}}]{Curty2014}%
  \BibitemOpen
  \bibfield  {author} {\bibinfo {author} {\bibfnamefont {M.}~\bibnamefont
  {Curty}}, \bibinfo {author} {\bibfnamefont {F.}~\bibnamefont {Xu}}, \bibinfo
  {author} {\bibfnamefont {C.~C.~W.}\ \bibnamefont {Lim}}, \bibinfo {author}
  {\bibfnamefont {K.}~\bibnamefont {Tamaki}}, \ and\ \bibinfo {author}
  {\bibfnamefont {H.-K.}\ \bibnamefont {Lo}},\ }\href {\doibase
  10.1038/ncomms4732} {\bibfield  {journal} {\bibinfo  {journal} {Nat.
  Commun.}\ }\textbf {\bibinfo {volume} {5}},\ \bibinfo {pages} {3732}
  (\bibinfo {year} {2014})}\BibitemShut {NoStop}%
\bibitem [{\citenamefont {Kobayashi}\ \emph {et~al.}(2014)\citenamefont
  {Kobayashi}, \citenamefont {Tomita},\ and\ \citenamefont
  {Okamoto}}]{kobayashi_evaluation_2014}%
  \BibitemOpen
  \bibfield  {author} {\bibinfo {author} {\bibfnamefont {T.}~\bibnamefont
  {Kobayashi}}, \bibinfo {author} {\bibfnamefont {A.}~\bibnamefont {Tomita}}, \
  and\ \bibinfo {author} {\bibfnamefont {A.}~\bibnamefont {Okamoto}},\
  }\href@noop {} {\bibfield  {journal} {\bibinfo  {journal} {Phys. Rev. A}\
  }\textbf {\bibinfo {volume} {90}},\ \bibinfo {pages} {032320} (\bibinfo
  {year} {2014})}\BibitemShut {NoStop}%
\bibitem [{\citenamefont {Yuan}\ \emph
  {et~al.}(2014{\natexlab{a}})\citenamefont {Yuan}, \citenamefont {Lucamarini},
  \citenamefont {Dynes}, \citenamefont {Fröhlich}, \citenamefont {Plews},\
  and\ \citenamefont {Shields}}]{yuan_robust_2014}%
  \BibitemOpen
  \bibfield  {author} {\bibinfo {author} {\bibfnamefont {Z.~L.}\ \bibnamefont
  {Yuan}}, \bibinfo {author} {\bibfnamefont {M.}~\bibnamefont {Lucamarini}},
  \bibinfo {author} {\bibfnamefont {J.~F.}\ \bibnamefont {Dynes}}, \bibinfo
  {author} {\bibfnamefont {B.}~\bibnamefont {Fröhlich}}, \bibinfo {author}
  {\bibfnamefont {A.}~\bibnamefont {Plews}}, \ and\ \bibinfo {author}
  {\bibfnamefont {A.~J.}\ \bibnamefont {Shields}},\ }\href@noop {} {\bibfield
  {journal} {\bibinfo  {journal} {Appl. Phys. Lett.}\ }\textbf {\bibinfo
  {volume} {104}},\ \bibinfo {pages} {261112} (\bibinfo {year}
  {2014}{\natexlab{a}})}\BibitemShut {NoStop}%
\bibitem [{\citenamefont {Tang}\ \emph {et~al.}(2013)\citenamefont {Tang},
  \citenamefont {Yin}, \citenamefont {Ma}, \citenamefont {Fung}, \citenamefont
  {Liu}, \citenamefont {Yong}, \citenamefont {Chen}, \citenamefont {Peng},
  \citenamefont {Chen},\ and\ \citenamefont {Pan}}]{tang_source_2013}%
  \BibitemOpen
  \bibfield  {author} {\bibinfo {author} {\bibfnamefont {Y.-L.}\ \bibnamefont
  {Tang}}, \bibinfo {author} {\bibfnamefont {H.-L.}\ \bibnamefont {Yin}},
  \bibinfo {author} {\bibfnamefont {X.}~\bibnamefont {Ma}}, \bibinfo {author}
  {\bibfnamefont {C.-H.~F.}\ \bibnamefont {Fung}}, \bibinfo {author}
  {\bibfnamefont {Y.}~\bibnamefont {Liu}}, \bibinfo {author} {\bibfnamefont
  {H.-L.}\ \bibnamefont {Yong}}, \bibinfo {author} {\bibfnamefont {T.-Y.}\
  \bibnamefont {Chen}}, \bibinfo {author} {\bibfnamefont {C.-Z.}\ \bibnamefont
  {Peng}}, \bibinfo {author} {\bibfnamefont {Z.-B.}\ \bibnamefont {Chen}}, \
  and\ \bibinfo {author} {\bibfnamefont {J.-W.}\ \bibnamefont {Pan}},\
  }\href@noop {} {\bibfield  {journal} {\bibinfo  {journal} {Phys. Rev. A}\
  }\textbf {\bibinfo {volume} {88}},\ \bibinfo {pages} {022308} (\bibinfo
  {year} {2013})}\BibitemShut {NoStop}%
\bibitem [{\citenamefont {Zhang}\ \emph {et~al.}(2021)\citenamefont {Zhang},
  \citenamefont {Primaatmaja}, \citenamefont {Haw}, \citenamefont {Gong},
  \citenamefont {Wang},\ and\ \citenamefont {Lim}}]{zhang_securing_2020}%
  \BibitemOpen
  \bibfield  {author} {\bibinfo {author} {\bibfnamefont {G.}~\bibnamefont
  {Zhang}}, \bibinfo {author} {\bibfnamefont {I.~W.}\ \bibnamefont
  {Primaatmaja}}, \bibinfo {author} {\bibfnamefont {J.~Y.}\ \bibnamefont
  {Haw}}, \bibinfo {author} {\bibfnamefont {X.}~\bibnamefont {Gong}}, \bibinfo
  {author} {\bibfnamefont {C.}~\bibnamefont {Wang}}, \ and\ \bibinfo {author}
  {\bibfnamefont {C.~C.~W.}\ \bibnamefont {Lim}},\ }\href@noop {} {\bibfield
  {journal} {\bibinfo  {journal} {PRX Quantum}\ }\textbf {\bibinfo {volume}
  {2}},\ \bibinfo {pages} {030304} (\bibinfo {year} {2021})}\BibitemShut
  {NoStop}%
\bibitem [{\citenamefont {Lucamarini}\ \emph {et~al.}(2015)\citenamefont
  {Lucamarini}, \citenamefont {Choi}, \citenamefont {Ward}, \citenamefont
  {Dynes}, \citenamefont {Yuan},\ and\ \citenamefont
  {Shields}}]{lucamarini_practical_2015}%
  \BibitemOpen
  \bibfield  {author} {\bibinfo {author} {\bibfnamefont {M.}~\bibnamefont
  {Lucamarini}}, \bibinfo {author} {\bibfnamefont {I.}~\bibnamefont {Choi}},
  \bibinfo {author} {\bibfnamefont {M.}~\bibnamefont {Ward}}, \bibinfo {author}
  {\bibfnamefont {J.}~\bibnamefont {Dynes}}, \bibinfo {author} {\bibfnamefont
  {Z.}~\bibnamefont {Yuan}}, \ and\ \bibinfo {author} {\bibfnamefont
  {A.}~\bibnamefont {Shields}},\ }\href@noop {} {\bibfield  {journal} {\bibinfo
   {journal} {Phys. Rev. X}\ }\textbf {\bibinfo {volume} {5}},\ \bibinfo
  {pages} {031030} (\bibinfo {year} {2015})}\BibitemShut {NoStop}%
\bibitem [{ISO(2011)}]{ISO_Fingerprint}%
  \BibitemOpen
  \href@noop {} {\emph {\bibinfo {title} {Information technology —
  {Biometric} data interchange formats — {Part} 2: {Finger} minutiae
  data}}},\ \bibinfo {type} {{ISO}/{IEC}}\ \bibinfo {number} {19794-2:2011}\
  (\bibinfo  {institution} {International Organization for Standardization},\
  \bibinfo {address} {Geneva, Switzerland},\ \bibinfo {year}
  {2011})\BibitemShut {NoStop}%
\bibitem [{\citenamefont {Maltoni}\ \emph {et~al.}(2009)\citenamefont
  {Maltoni}, \citenamefont {Maio}, \citenamefont {Jain},\ and\ \citenamefont
  {Prabhakar}}]{Maltoni2009}%
  \BibitemOpen
  \bibfield  {author} {\bibinfo {author} {\bibfnamefont {D.}~\bibnamefont
  {Maltoni}}, \bibinfo {author} {\bibfnamefont {D.}~\bibnamefont {Maio}},
  \bibinfo {author} {\bibfnamefont {A.~K.}\ \bibnamefont {Jain}}, \ and\
  \bibinfo {author} {\bibfnamefont {S.}~\bibnamefont {Prabhakar}},\ }\href
  {\doibase 10.1007/978-1-84882-254-2} {\emph {\bibinfo {title} {Handbook of
  {Fingerprint} {Recognition}}}},\ \bibinfo {edition} {2nd}\ ed.\ (\bibinfo
  {publisher} {Springer-Verlag},\ \bibinfo {address} {London},\ \bibinfo {year}
  {2009})\BibitemShut {NoStop}%
\bibitem [{\citenamefont {Kayaoglu}\ \emph {et~al.}(2013)\citenamefont
  {Kayaoglu}, \citenamefont {Topcu},\ and\ \citenamefont
  {Uludag}}]{Kayaoglu2013}%
  \BibitemOpen
  \bibfield  {author} {\bibinfo {author} {\bibfnamefont {M.}~\bibnamefont
  {Kayaoglu}}, \bibinfo {author} {\bibfnamefont {B.}~\bibnamefont {Topcu}}, \
  and\ \bibinfo {author} {\bibfnamefont {U.}~\bibnamefont {Uludag}},\ }\href
  {http://arxiv.org/abs/1305.1443} {\bibfield  {journal} {\bibinfo  {journal}
  {arXiv:1305.1443 [cs]}\ } (\bibinfo {year} {2013})}\BibitemShut {NoStop}%
\bibitem [{\citenamefont {Yuan}\ \emph
  {et~al.}(2014{\natexlab{b}})\citenamefont {Yuan}, \citenamefont {Lucamarini},
  \citenamefont {Dynes}, \citenamefont {Fröhlich}, \citenamefont {Ward},\ and\
  \citenamefont {Shields}}]{yuan_interference_2014}%
  \BibitemOpen
  \bibfield  {author} {\bibinfo {author} {\bibfnamefont {Z.}~\bibnamefont
  {Yuan}}, \bibinfo {author} {\bibfnamefont {M.}~\bibnamefont {Lucamarini}},
  \bibinfo {author} {\bibfnamefont {J.}~\bibnamefont {Dynes}}, \bibinfo
  {author} {\bibfnamefont {B.}~\bibnamefont {Fröhlich}}, \bibinfo {author}
  {\bibfnamefont {M.}~\bibnamefont {Ward}}, \ and\ \bibinfo {author}
  {\bibfnamefont {A.}~\bibnamefont {Shields}},\ }\href@noop {} {\bibfield
  {journal} {\bibinfo  {journal} {Phys. Rev. Applied}\ }\textbf {\bibinfo
  {volume} {2}},\ \bibinfo {pages} {064006} (\bibinfo {year}
  {2014}{\natexlab{b}})}\BibitemShut {NoStop}%
\bibitem [{\citenamefont {Ma}\ and\ \citenamefont
  {Razavi}(2012)}]{ma_alternative_2012}%
  \BibitemOpen
  \bibfield  {author} {\bibinfo {author} {\bibfnamefont {X.}~\bibnamefont
  {Ma}}\ and\ \bibinfo {author} {\bibfnamefont {M.}~\bibnamefont {Razavi}},\
  }\href@noop {} {\bibfield  {journal} {\bibinfo  {journal} {Phys. Rev. A}\
  }\textbf {\bibinfo {volume} {86}},\ \bibinfo {pages} {062319} (\bibinfo
  {year} {2012})}\BibitemShut {NoStop}%
\bibitem [{\citenamefont {Wang}\ \emph {et~al.}(2017)\citenamefont {Wang},
  \citenamefont {Wang}, \citenamefont {Chen}, \citenamefont {Wang},
  \citenamefont {Chen}, \citenamefont {Yin}, \citenamefont {He}, \citenamefont
  {Guo},\ and\ \citenamefont {Han}}]{wang_realistic_2017}%
  \BibitemOpen
  \bibfield  {author} {\bibinfo {author} {\bibfnamefont {C.}~\bibnamefont
  {Wang}}, \bibinfo {author} {\bibfnamefont {F.~X.}\ \bibnamefont {Wang}},
  \bibinfo {author} {\bibfnamefont {H.}~\bibnamefont {Chen}}, \bibinfo {author}
  {\bibfnamefont {S.}~\bibnamefont {Wang}}, \bibinfo {author} {\bibfnamefont
  {W.}~\bibnamefont {Chen}}, \bibinfo {author} {\bibfnamefont {Z.~Q.}\
  \bibnamefont {Yin}}, \bibinfo {author} {\bibfnamefont {D.~Y.}\ \bibnamefont
  {He}}, \bibinfo {author} {\bibfnamefont {G.~C.}\ \bibnamefont {Guo}}, \ and\
  \bibinfo {author} {\bibfnamefont {Z.~F.}\ \bibnamefont {Han}},\ }\href@noop
  {} {\bibfield  {journal} {\bibinfo  {journal} {J. Lightwave Technol.}\
  }\textbf {\bibinfo {volume} {35}},\ \bibinfo {pages} {4996} (\bibinfo {year}
  {2017})}\BibitemShut {NoStop}%
\bibitem [{\citenamefont {Wei}\ \emph {et~al.}(2020)\citenamefont {Wei},
  \citenamefont {Li}, \citenamefont {Tan}, \citenamefont {Li}, \citenamefont
  {Min}, \citenamefont {Zhang}, \citenamefont {Li}, \citenamefont {You},
  \citenamefont {Wang}, \citenamefont {Jiang}, \citenamefont {Chen},
  \citenamefont {Liao}, \citenamefont {Peng}, \citenamefont {Xu},\ and\
  \citenamefont {Pan}}]{wei_high-speed_2020}%
  \BibitemOpen
  \bibfield  {author} {\bibinfo {author} {\bibfnamefont {K.}~\bibnamefont
  {Wei}}, \bibinfo {author} {\bibfnamefont {W.}~\bibnamefont {Li}}, \bibinfo
  {author} {\bibfnamefont {H.}~\bibnamefont {Tan}}, \bibinfo {author}
  {\bibfnamefont {Y.}~\bibnamefont {Li}}, \bibinfo {author} {\bibfnamefont
  {H.}~\bibnamefont {Min}}, \bibinfo {author} {\bibfnamefont {W.-J.}\
  \bibnamefont {Zhang}}, \bibinfo {author} {\bibfnamefont {H.}~\bibnamefont
  {Li}}, \bibinfo {author} {\bibfnamefont {L.}~\bibnamefont {You}}, \bibinfo
  {author} {\bibfnamefont {Z.}~\bibnamefont {Wang}}, \bibinfo {author}
  {\bibfnamefont {X.}~\bibnamefont {Jiang}}, \bibinfo {author} {\bibfnamefont
  {T.-Y.}\ \bibnamefont {Chen}}, \bibinfo {author} {\bibfnamefont {S.-K.}\
  \bibnamefont {Liao}}, \bibinfo {author} {\bibfnamefont {C.-Z.}\ \bibnamefont
  {Peng}}, \bibinfo {author} {\bibfnamefont {F.}~\bibnamefont {Xu}}, \ and\
  \bibinfo {author} {\bibfnamefont {J.-W.}\ \bibnamefont {Pan}},\ }\href@noop
  {} {\bibfield  {journal} {\bibinfo  {journal} {Phys. Rev. X}\ }\textbf
  {\bibinfo {volume} {10}},\ \bibinfo {pages} {031030} (\bibinfo {year}
  {2020})}\BibitemShut {NoStop}%
\bibitem [{\citenamefont {Dauler}\ \emph {et~al.}(2006)\citenamefont {Dauler},
  \citenamefont {Robinson}, \citenamefont {Kerman}, \citenamefont {Anant},
  \citenamefont {Barron}, \citenamefont {Berggren}, \citenamefont {Caplan},
  \citenamefont {Carney}, \citenamefont {Hamilton}, \citenamefont {Rosfjord},
  \citenamefont {Stevens},\ and\ \citenamefont {Yang}}]{dauler_125-gbits_2006}%
  \BibitemOpen
  \bibfield  {author} {\bibinfo {author} {\bibfnamefont {E.~A.}\ \bibnamefont
  {Dauler}}, \bibinfo {author} {\bibfnamefont {B.~S.}\ \bibnamefont
  {Robinson}}, \bibinfo {author} {\bibfnamefont {A.~J.}\ \bibnamefont
  {Kerman}}, \bibinfo {author} {\bibfnamefont {V.}~\bibnamefont {Anant}},
  \bibinfo {author} {\bibfnamefont {R.~J.}\ \bibnamefont {Barron}}, \bibinfo
  {author} {\bibfnamefont {K.~K.}\ \bibnamefont {Berggren}}, \bibinfo {author}
  {\bibfnamefont {D.~O.}\ \bibnamefont {Caplan}}, \bibinfo {author}
  {\bibfnamefont {J.~J.}\ \bibnamefont {Carney}}, \bibinfo {author}
  {\bibfnamefont {S.~A.}\ \bibnamefont {Hamilton}}, \bibinfo {author}
  {\bibfnamefont {K.~M.}\ \bibnamefont {Rosfjord}}, \bibinfo {author}
  {\bibfnamefont {M.~L.}\ \bibnamefont {Stevens}}, \ and\ \bibinfo {author}
  {\bibfnamefont {J.~K.~W.}\ \bibnamefont {Yang}},\ }in\ \href@noop {} {\emph
  {\bibinfo {booktitle} {Advanced {Photon} {Counting} {Techniques}}}},\ Vol.\
  \bibinfo {volume} {6372}\ (\bibinfo  {publisher} {SPIE},\ \bibinfo {year}
  {2006})\ pp.\ \bibinfo {pages} {286 -- 293}\BibitemShut {NoStop}%
\bibitem [{\citenamefont {Chen}\ \emph {et~al.}(2021)\citenamefont {Chen},
  \citenamefont {Zhang}, \citenamefont {Liu}, \citenamefont {Jiang},
  \citenamefont {Zhang}, \citenamefont {Han}, \citenamefont {Ma}, \citenamefont
  {Hu}, \citenamefont {Li}, \citenamefont {Liu}, \citenamefont {Zhou},
  \citenamefont {Jiang}, \citenamefont {Chen}, \citenamefont {Li},
  \citenamefont {You}, \citenamefont {Wang}, \citenamefont {Wang},
  \citenamefont {Zhang},\ and\ \citenamefont {Pan}}]{chen_twin-field_2021}%
  \BibitemOpen
  \bibfield  {author} {\bibinfo {author} {\bibfnamefont {J.-P.}\ \bibnamefont
  {Chen}}, \bibinfo {author} {\bibfnamefont {C.}~\bibnamefont {Zhang}},
  \bibinfo {author} {\bibfnamefont {Y.}~\bibnamefont {Liu}}, \bibinfo {author}
  {\bibfnamefont {C.}~\bibnamefont {Jiang}}, \bibinfo {author} {\bibfnamefont
  {W.-J.}\ \bibnamefont {Zhang}}, \bibinfo {author} {\bibfnamefont {Z.-Y.}\
  \bibnamefont {Han}}, \bibinfo {author} {\bibfnamefont {S.-Z.}\ \bibnamefont
  {Ma}}, \bibinfo {author} {\bibfnamefont {X.-L.}\ \bibnamefont {Hu}}, \bibinfo
  {author} {\bibfnamefont {Y.-H.}\ \bibnamefont {Li}}, \bibinfo {author}
  {\bibfnamefont {H.}~\bibnamefont {Liu}}, \bibinfo {author} {\bibfnamefont
  {F.}~\bibnamefont {Zhou}}, \bibinfo {author} {\bibfnamefont {H.-F.}\
  \bibnamefont {Jiang}}, \bibinfo {author} {\bibfnamefont {T.-Y.}\ \bibnamefont
  {Chen}}, \bibinfo {author} {\bibfnamefont {H.}~\bibnamefont {Li}}, \bibinfo
  {author} {\bibfnamefont {L.-X.}\ \bibnamefont {You}}, \bibinfo {author}
  {\bibfnamefont {Z.}~\bibnamefont {Wang}}, \bibinfo {author} {\bibfnamefont
  {X.-B.}\ \bibnamefont {Wang}}, \bibinfo {author} {\bibfnamefont
  {Q.}~\bibnamefont {Zhang}}, \ and\ \bibinfo {author} {\bibfnamefont {J.-W.}\
  \bibnamefont {Pan}},\ }\href@noop {} {\bibfield  {journal} {\bibinfo
  {journal} {Nat. Photon.}\ }\textbf {\bibinfo {volume} {15}},\ \bibinfo
  {pages} {570} (\bibinfo {year} {2021})}\BibitemShut {NoStop}%
\bibitem [{\citenamefont {Kiktenko}\ \emph {et~al.}(2017)\citenamefont
  {Kiktenko}, \citenamefont {Trushechkin}, \citenamefont {Lim}, \citenamefont
  {Kurochkin},\ and\ \citenamefont {Fedorov}}]{kiktenko_symmetric_2017}%
  \BibitemOpen
  \bibfield  {author} {\bibinfo {author} {\bibfnamefont {E.}~\bibnamefont
  {Kiktenko}}, \bibinfo {author} {\bibfnamefont {A.}~\bibnamefont
  {Trushechkin}}, \bibinfo {author} {\bibfnamefont {C.}~\bibnamefont {Lim}},
  \bibinfo {author} {\bibfnamefont {Y.}~\bibnamefont {Kurochkin}}, \ and\
  \bibinfo {author} {\bibfnamefont {A.}~\bibnamefont {Fedorov}},\ }\href@noop
  {} {\bibfield  {journal} {\bibinfo  {journal} {Phys. Rev. Applied}\ }\textbf
  {\bibinfo {volume} {8}},\ \bibinfo {pages} {044017} (\bibinfo {year}
  {2017})}\BibitemShut {NoStop}%
\bibitem [{\citenamefont {Elkouss}\ \emph {et~al.}(2009)\citenamefont
  {Elkouss}, \citenamefont {Leverrier}, \citenamefont {All{\'e}aume},\ and\
  \citenamefont {Boutros}}]{elkouss_efficient_2009}%
  \BibitemOpen
  \bibfield  {author} {\bibinfo {author} {\bibfnamefont {D.}~\bibnamefont
  {Elkouss}}, \bibinfo {author} {\bibfnamefont {A.}~\bibnamefont {Leverrier}},
  \bibinfo {author} {\bibfnamefont {R.}~\bibnamefont {All{\'e}aume}}, \ and\
  \bibinfo {author} {\bibfnamefont {J.}~\bibnamefont {Boutros}},\ }\href@noop
  {} {\bibfield  {journal} {\bibinfo  {journal} {2009 IEEE International
  Symposium on Information Theory}\ ,\ \bibinfo {pages} {1879}} (\bibinfo
  {year} {2009})}\BibitemShut {NoStop}%
\bibitem [{\citenamefont {Tomamichel}\ \emph {et~al.}(2017)\citenamefont
  {Tomamichel}, \citenamefont {Martinez-Mateo}, \citenamefont {Pacher},\ and\
  \citenamefont {Elkouss}}]{tomamichel_fundamental_2017}%
  \BibitemOpen
  \bibfield  {author} {\bibinfo {author} {\bibfnamefont {M.}~\bibnamefont
  {Tomamichel}}, \bibinfo {author} {\bibfnamefont {J.}~\bibnamefont
  {Martinez-Mateo}}, \bibinfo {author} {\bibfnamefont {C.}~\bibnamefont
  {Pacher}}, \ and\ \bibinfo {author} {\bibfnamefont {D.}~\bibnamefont
  {Elkouss}},\ }\href@noop {} {\bibfield  {journal} {\bibinfo  {journal}
  {Quantum Inf Process}\ }\textbf {\bibinfo {volume} {16}},\ \bibinfo {pages}
  {280} (\bibinfo {year} {2017})}\BibitemShut {NoStop}%
\bibitem [{\citenamefont {Krawczyk}(1994)}]{krawczyk_lfsr-based_1994}%
  \BibitemOpen
  \bibfield  {author} {\bibinfo {author} {\bibfnamefont {H.}~\bibnamefont
  {Krawczyk}},\ }\href@noop {} {\bibfield  {journal} {\bibinfo  {journal}
  {Advances in Cryptology-CRYPTO'94}\ }\textbf {\bibinfo {volume} {839}},\
  \bibinfo {pages} {129} (\bibinfo {year} {1994})}\BibitemShut {NoStop}%
\bibitem [{\citenamefont
  {Robling~Denning}(1982)}]{robling_denning_cryptography_1982}%
  \BibitemOpen
  \bibfield  {author} {\bibinfo {author} {\bibfnamefont {D.~E.}\ \bibnamefont
  {Robling~Denning}},\ }\href@noop {} {\emph {\bibinfo {title} {Cryptography
  and {Data} {Security}}}}\ (\bibinfo  {publisher} {Addison-Wesley Longman
  Publishing Co., Inc.},\ \bibinfo {year} {1982})\BibitemShut {NoStop}%
\bibitem [{\citenamefont {Kim}\ and\ \citenamefont
  {Solomon}(2016)}]{kim_fundamentals_2016}%
  \BibitemOpen
  \bibfield  {author} {\bibinfo {author} {\bibfnamefont {D.}~\bibnamefont
  {Kim}}\ and\ \bibinfo {author} {\bibfnamefont {M.}~\bibnamefont {Solomon}},\
  }\href@noop {} {\emph {\bibinfo {title} {Fundamentals of {Information}
  {Systems} {Security}}}},\ \bibinfo {edition} {3rd}\ ed.\ (\bibinfo
  {publisher} {JONES \& BARTLETT PUB Incorporated},\ \bibinfo {year}
  {2016})\BibitemShut {NoStop}%
\bibitem [{\citenamefont {Sun}\ \emph {et~al.}(2015)\citenamefont {Sun},
  \citenamefont {Xu}, \citenamefont {Jiang}, \citenamefont {Ma}, \citenamefont
  {Lo},\ and\ \citenamefont {Liang}}]{sun_effect_2015}%
  \BibitemOpen
  \bibfield  {author} {\bibinfo {author} {\bibfnamefont {S.-H.}\ \bibnamefont
  {Sun}}, \bibinfo {author} {\bibfnamefont {F.}~\bibnamefont {Xu}}, \bibinfo
  {author} {\bibfnamefont {M.-S.}\ \bibnamefont {Jiang}}, \bibinfo {author}
  {\bibfnamefont {X.-C.}\ \bibnamefont {Ma}}, \bibinfo {author} {\bibfnamefont
  {H.-K.}\ \bibnamefont {Lo}}, \ and\ \bibinfo {author} {\bibfnamefont {L.-M.}\
  \bibnamefont {Liang}},\ }\href@noop {} {\bibfield  {journal} {\bibinfo
  {journal} {Phys. Rev. A}\ }\textbf {\bibinfo {volume} {92}},\ \bibinfo
  {pages} {022304} (\bibinfo {year} {2015})}\BibitemShut {NoStop}%
\bibitem [{\citenamefont {Huang}\ \emph {et~al.}(2019)\citenamefont {Huang},
  \citenamefont {Navarrete}, \citenamefont {Sun}, \citenamefont {Chaiwongkhot},
  \citenamefont {Curty},\ and\ \citenamefont
  {Makarov}}]{huang_laser-seeding_2019}%
  \BibitemOpen
  \bibfield  {author} {\bibinfo {author} {\bibfnamefont {A.}~\bibnamefont
  {Huang}}, \bibinfo {author} {\bibfnamefont {{\'A}.}~\bibnamefont
  {Navarrete}}, \bibinfo {author} {\bibfnamefont {S.-H.}\ \bibnamefont {Sun}},
  \bibinfo {author} {\bibfnamefont {P.}~\bibnamefont {Chaiwongkhot}}, \bibinfo
  {author} {\bibfnamefont {M.}~\bibnamefont {Curty}}, \ and\ \bibinfo {author}
  {\bibfnamefont {V.}~\bibnamefont {Makarov}},\ }\href@noop {} {\bibfield
  {journal} {\bibinfo  {journal} {Phys. Rev. Applied}\ }\textbf {\bibinfo
  {volume} {12}},\ \bibinfo {pages} {064043} (\bibinfo {year}
  {2019})}\BibitemShut {NoStop}%
\bibitem [{\citenamefont {Huang}\ \emph {et~al.}(2020)\citenamefont {Huang},
  \citenamefont {Li}, \citenamefont {Egorov}, \citenamefont {Tchouragoulov},
  \citenamefont {Kumar},\ and\ \citenamefont
  {Makarov}}]{huang_laser-damage_2020}%
  \BibitemOpen
  \bibfield  {author} {\bibinfo {author} {\bibfnamefont {A.}~\bibnamefont
  {Huang}}, \bibinfo {author} {\bibfnamefont {R.}~\bibnamefont {Li}}, \bibinfo
  {author} {\bibfnamefont {V.}~\bibnamefont {Egorov}}, \bibinfo {author}
  {\bibfnamefont {S.}~\bibnamefont {Tchouragoulov}}, \bibinfo {author}
  {\bibfnamefont {K.}~\bibnamefont {Kumar}}, \ and\ \bibinfo {author}
  {\bibfnamefont {V.}~\bibnamefont {Makarov}},\ }\href@noop {} {\bibfield
  {journal} {\bibinfo  {journal} {Phys. Rev. Applied}\ }\textbf {\bibinfo
  {volume} {13}},\ \bibinfo {pages} {034017} (\bibinfo {year}
  {2020})}\BibitemShut {NoStop}%
\end{thebibliography}%


%

\end{document}